\def\year{2020}\relax
\documentclass[letterpaper]{article} 
\usepackage{aaai20}  
\usepackage{times}  
\usepackage{helvet} 
\usepackage{courier}  
\usepackage[hyphens]{url}  
\usepackage{graphicx} 
\usepackage{graphicx}  
\usepackage{amsmath}
\usepackage{multirow}
\usepackage{booktabs}
\usepackage{subcaption}
\usepackage{adjustbox}
\title{Look, Read and Feel: Benchmarking Ads Understanding with Multimodal Multitask Learning}

\author{\Large \textbf{Huaizheng Zhang, Yong Luo, Qiming Ai, Yonggang Wen}\\ 
\textsuperscript{}Nanyang Technological University\\ 
\{huaizhen001, yluo, qmai, ygwen\}@ntu.edu.sg.
}
\begin{document}

\maketitle

\begin{abstract}
Given the massive market of advertising and the sharply increasing online multimedia content (such as videos), it is now fashionable to promote advertisements (ads) together with the multimedia content. It is exhausted to find relevant ads to match the provided content manually, and hence, some automatic advertising techniques are developed. Since ads are usually hard to understand only according to its visual appearance due to the contained visual metaphor, some other modalities, such as the contained texts, should be exploited for understanding. To further improve user experience, it is necessary to understand both the topic and sentiment of the ads. This motivates us to develop a novel deep multimodal multitask framework to integrate multiple modalities to achieve effective topic and sentiment prediction simultaneously for ads understanding. In particular, our model first extracts multimodal information from ads and learn high-level and comparable representations. The visual metaphor of the ad is decoded in an unsupervised manner. The obtained representations are then fed into the proposed hierarchical multimodal attention modules to learn task-specific representations for final prediction. A multitask loss function is also designed to train both the topic and sentiment prediction models jointly in an end-to-end manner. We conduct extensive experiments on the latest and large advertisement dataset and achieve state-of-the-art performance for both prediction tasks. The obtained results could be utilized as a benchmark for ads understanding.
\end{abstract}

\section{Introduction}
Advertising has a pivotal role in the global economy and the revenue of numerous companies. For instance, it is predicted that Google's advertising revenues will improve about 15\% to \$39.92 billion in 2018, and the improvement is 17\% to \$21 billion for Facebook \cite{googleads}. Since there is tremendous multimedia content (such as videos and TV shows) on the web, it is now fashionable to promote the ads together with multimedia content. Manually selecting an ad to match a provided multimedia content is time-consuming and labor-intensive. Thus some automatic advertising techniques are developed, such as the contextual advertising \cite{mei2007videosense} method, which aims to find the most relevant ad to a provided content without annoying customers. Therefore, it is necessary to understand both the multimedia content and ads thoroughly.


\begin{figure}[t]
\centering
\includegraphics[width=1.0\columnwidth]{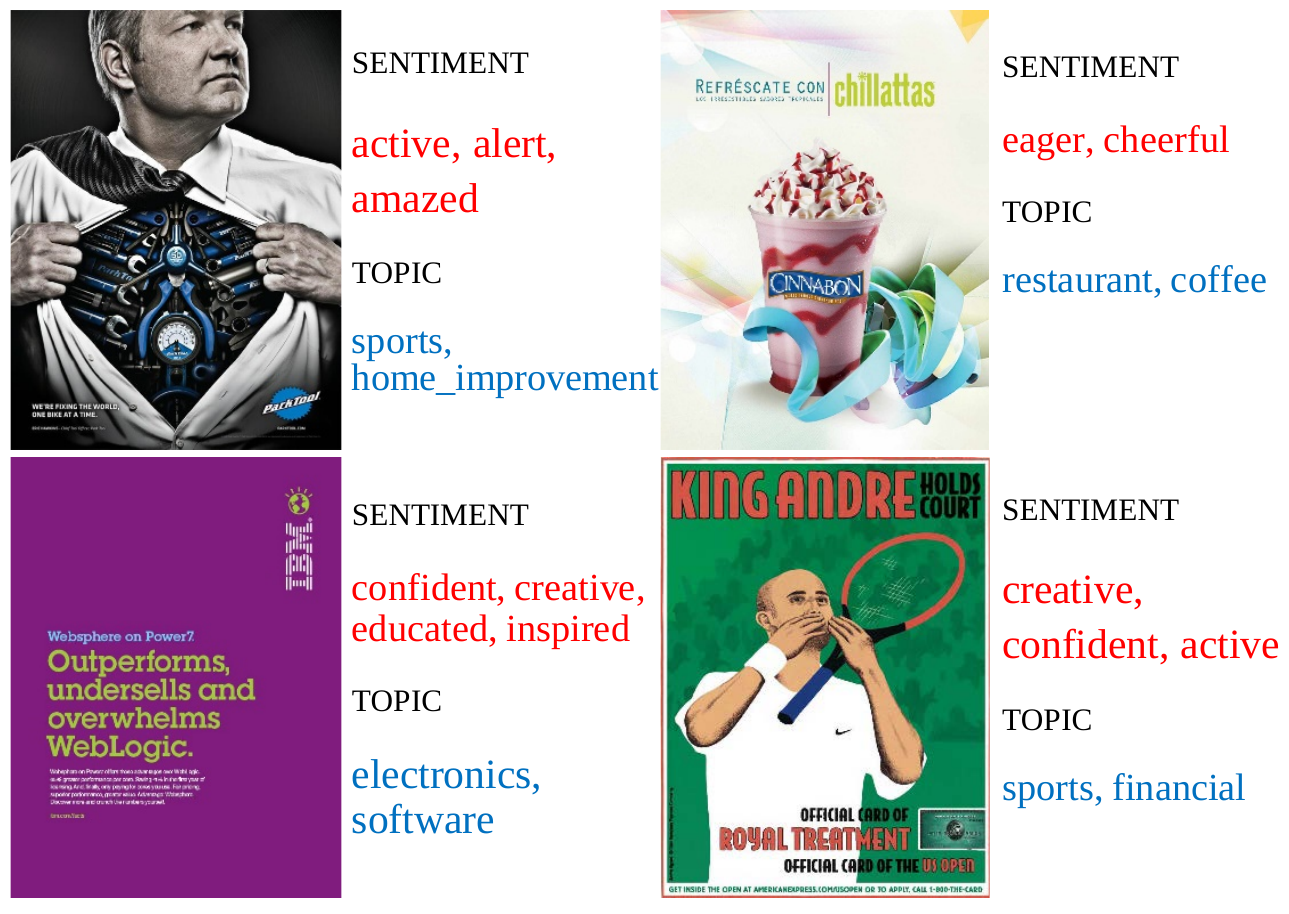} 
\caption{Ad examples. Visual rhetoric exists widely in ads design. Besides, different from natural images, ad images contains multimodal information such as visual image and associated texts, which can be utilized for better understanding of ads.}
\label{fig:/ads_example}
\end{figure}

Benefiting from the deep learning technique, typical multimedia content analysis has achieved excellent progress \cite{vedula2017multimodal,madhok2018semantic}. However, there is much less success in understanding ads, which are often much harder to understanding since they usually contain much visual \textit{rhetoric} \cite{hussain2017automatic} to attract, communicate with, and even persuade customers. Some example ad images are shown in Figure~\ref{fig:/ads_example}, where we can see that it is often hard to understand an ad only according to its visual appearance (\emph{look}). Fortunately, the contained or associated texts can usually indicate some underlying implication. Therefore, it is beneficial to integrate both the visual and textual information (\emph{read}) for effective ad understanding. In addition, to improve the user experience when promoting the ads, we should not only understand its topic, but also the emotion (\emph{feel}) it conveys. For example, for the same living room scene in a TV show, the atmosphere may vary due to the difference of characters' situations, and it is inappropriate to insert an ad that conveys ``\textit{creative}'' or ``\textit{inspired}'' to the scene where characters are painful. Consequently, it is necessary to predict both the topic and sentiment of an ad for non-intrusive user experience. Since both the topic and sentiment are related to the same ad, we propose to learn the two prediction models simultaneously to enable them to help each other in training.

Based on these considerations, we propose a novel deep multimodal multitask framework for ad understanding, where different types of information are integrated to predict both the topic and sentiment of an ad simultaneously. To our best knowledge, this is the first framework that unifies the topic and sentiment understanding of ads. In particular, we first extract different types of information, such as objects and contained texts from the ad using some existing techniques, such as the pre-trained object or image representation models and OCR \cite{smith2007overview}. To recognize and understand the visual \textit{rhetoric}, autoencoder is introduced to decode the object representation in an unsupervised manner. For the whole image and extracted text representations, multi-layer perception (MLP) and BLSTM \cite{liu2016agreement} are added to learn high-level and comparable representations. The parameters in these modules are shared by different tasks, and thus the total number of parameters can be reduced significantly. The obtained representations are then fed into different sub-networks, where a novel hierarchical multimodal attention module is designed to capture both the intra-modal and inter-modal importance for different tasks. Finally, a multitask loss function is designed to minimize the topic and sentiment prediction, as well as the decoding reconstruction losses simultaneously.

There exist some preliminary attempts for automatic advertising. For example, \cite{mei2007videosense} proposed a framework to insert ads into videos based on global textual relevance gathered from video meta data and local visual-aural relevance gathered from low-level image and audio features. A similar system is presented in \cite{xiang2015salad}, but more advanced deep learning technique is introduced to analyze both video content and ad images. Different from these topic-only analysis works, \cite{vedula2017multimodal} developed an ads recommendation system based on the sentiment of multimedia contents, and a unified framework to understand both topic and sentiment of the multimedia contents is developed in \cite{madhok2018semantic}. The main drawbacks of these works are: 1) the topic and sentiment analysis are only conducted for multimedia contents but not for ads; 2) the ads are processed in the same way as the common images/videos and the specific characteristics of ads are ignored. The proposed method is different and superior to these works in that: 1) both the topic and sentiment are analyzed for ads and the models are trained together to enable interaction between the two prediction tasks; 2) the multimodal nature of ads is fully exploited, and the feature importance is captured in a both intra-modal and inter-modal manner.

\begin{figure*}
\centering
\includegraphics[width=0.9\linewidth]{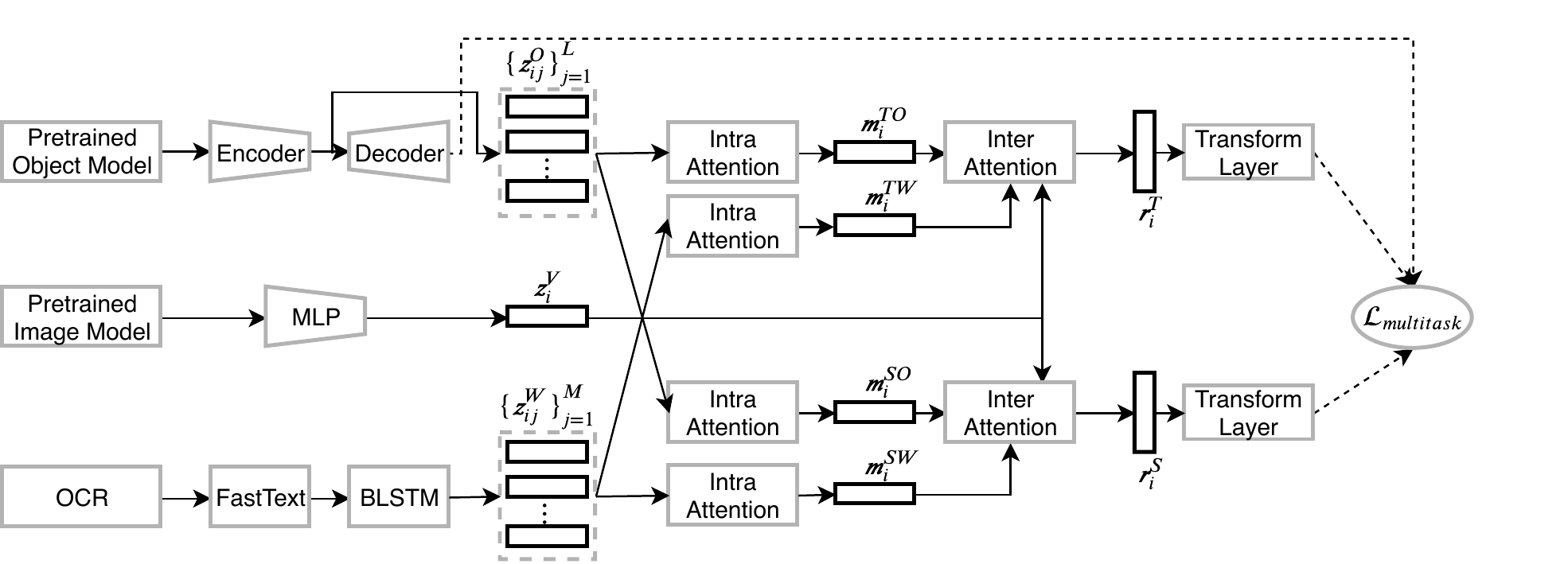} 
\caption{Network architecture. There are two main components in the proposed framework. Firstly, some existing models are adopted to extract object features, features of the whole ad, and the text features in the ad. The different types of features are passed through autoencoder, MLP and BLSTM respectively to learn high-level and comparable representations. Then the different representations are fused using the proposed hierarchical multimodal attention mechanism to learn task-specific representation for final prediction. The parameters in the first component are shared by different tasks, and the developed attention scheme is able to capture the feature importance in and between different modalities for different tasks.}
\label{fig:/framework}
\end{figure*}

We conduct extensive experiments to verify effectiveness of the proposed framework by comparing it with the ResNet baseline \cite{he2016deep} and some competitive or recently proposed multi-task/multi-label models  \cite{yeh2017learning,chen2019multi}. Improvements from 10\% to 78\% are achieved under the mean average precision (mAP) criterion. To summarize, our main contributions are:
\begin{itemize}
  \item The first multimodal multitask learning framework and a benchmark for ads understanding;
  \item A shared feature extraction module that makes full use of the multimodal information in ads and understand the visual \textit{rhetoric};
  \item A hierarchical multimodal attention module that effectively exploit the intra-modal and inter-modal information;
\end{itemize}





\section{Framework Design}

In this section, we first present the overview of the proposed architecture. We then divide the framework into three phrases and detail them. We finally describe how to train the framework.

\subsection{Architecture Overview}

\begin{align*}
    \text{\textbf{Input}: IMAGE} \rightarrow \text{\textbf{Output:}} \{(\boldsymbol{Y}^{T}_{i}, \boldsymbol{Y}^{S}_{i})\}
\end{align*}

We design a new neural network architecture (Figure \ref{fig:/framework}) to predict both topics and sentiments of ads in an end-to-end manner. Formally speaking, let $\mathcal{Y}^{T} = {\{y_{j}^{T}\}}^{q}_{j=1}$ be the topic label space with $q$ class labels and $\mathcal{Y}^{S} = {\{y_{k}^{S}\}}^{p}_{k=1}$ be the sentence label space with $p$ class labels. Given the training set $\mathcal{D} = \{(Image_{i}, \boldsymbol{Y}^{T}_{i}, \boldsymbol{Y}^{S}_{i}) | 1 \leq i \leq n  \}$, $\boldsymbol{Y}^{T}_{i} \subseteq \mathcal{Y}^{T}$ and $\boldsymbol{Y}^{S}_{i} \subseteq \mathcal{Y}^{S}$ are two sets of relevant topic and sentiment labels associated with $i$-th image. Our learned model $\mathcal{F}$ will assign multiple proper topic and sentiment labels to image ads.

During the inference stage, we divide the proposed framework into three phrases. In the first phrase, we use pretrained models and OCR to extract multi-modality features from ads and three kinds of sub-networks to learn shared multi-modality representations. These representation will be sent to the second phrase and processed by two hierarchical multimodal attention modules separately. The output from the second phrase is the task-specific representation. In the third phrase, we use a multitask prediction module to transform the learned representation and output final predictions.

\subsection{Shared Bottom Module}

\begin{align*}
    \text{\textbf{Input}: IMAGE} \rightarrow \text{\textbf{Output}: } \{(\boldsymbol{z}_{i}^{V}, {\{\boldsymbol{z}_{ij}^{O}\}}_{j=1}^{L}, {\{\boldsymbol{z}_{ij}^{W}\}}_{j=1}^{M})\}
\end{align*}
The first phrase of our framework is a shared bottom module as illustrated in Figure \ref{fig:/framework}. Given image ads, there are two main challenges need to be address in this phrase.
\begin{itemize}
    \item Since ads are designed to convey topics or emotions with multimodality information such as visual and text, how to separate these information and represent them is still open problem,
    \item As people use \textbf{\textit{rhetoric}} to decorate ads (Figure \ref{fig:/ads_example}), even the same object in one modality may have different meanings.
\end{itemize}


To tackle these challenges, we design a shared bottom module consisting of three components: a image component, an object component and a text component.

\textbf{Image component}. It is designed to extract a coarse global feature of an ad. There are two stages in this component which can be illustrated as follows:
\begin{gather}
    \boldsymbol{x}_{i}^{V} = f_{pooling}(f_{image}(\text{image}_{i}; \theta_{\text{fixed}})) \\
    \boldsymbol{z}_{i}^{V} = f_{MLP}(\boldsymbol{x}_{i}^{V};\theta)
    \label{eq:image_mlp}
\end{gather}
First, a pre-trained image classification model is adopted to transform the raw image into a feature map $\boldsymbol{S}^{C, H, W}$ where $C$ is the number of channels, $H$ is the height of the feature map and $W$ is the width of the feature map. Then a pooling technique (e.g., max or average) is applied to transform the feature map into a feature vector, $\boldsymbol{x}_i$. Second, we input the feature vector to a MLP to learn a more compact feature shared by the next two tasks. MLP can also align the length of the feature vectors to the other modality features.

\textbf{Object component}. To acquire regional features and decode visual metaphor, we rely on the our object component which includes two functions. A pre-trained object detection model (e.g. faster-rcnn)is applied to extract $L$ object-level features, $\{ \boldsymbol{x}_{i1}^{O}, \boldsymbol{x}_{i2}^{O}, ..., \boldsymbol{x}_{iL}^{O} \}$ from the bounding box proposals as follows:
\begin{align}
    {\{\boldsymbol{x}_{ij}^{O}\}}_{j=1}^{L} = f_{object}(\text{image}_i;\theta_{\text{fixed}})
    \label{eq:faster_rcnn}
\end{align}
The second function is to decode and represent the visual \textbf{\textit{rhetoric}} from these object features. To this end, we build a autoencoder as follows:
\begin{gather}
    \boldsymbol{z}_{ij}^{O} = f_{encoder}(\boldsymbol{x}_{ij}^{O}; \theta)\\
    \hat{\boldsymbol{x}}_{ij}^{O} = f_{decoder}(\boldsymbol{z}_{ij}^{O}; \theta)
    \label{eq:autoencoder}
\end{gather}
It contains an encoder and a decoder. The encoder projects the object features into a latent space where the features, $\{ \boldsymbol{z}_{i1}^{O}, \boldsymbol{z}_{i2}^{O}, ..., \boldsymbol{z}_{iL}^{O} \}$, have similar meanings are clustered together. The decoder reconstructs the latent features as $\{ \hat{\boldsymbol{x}}_{i1}^{O}, \hat{\boldsymbol{x}}_{i2}^{O}, ..., \hat{\boldsymbol{x}}_{iL}^{O} \}$ to make $\hat{\boldsymbol{x}}_{ij}^{O}$ similar to $\boldsymbol{x}_{ij}^{O}$. During the inference, we use the latent feature $\boldsymbol{z}_i$ as object representation.

Since visual \textbf{\textit{rhetoric}} has no supervised information, we train this part in an unsupervised manner with the reconstructed loss, $\mathcal{L}_{share}$.
\begin{align}
    \mathcal{L}_{share} = \frac{1}{n} \frac{1}{L} \sum_{i=1}^{n} \sum_{j=1}^{L} (\hat{\boldsymbol{x}}_{ij}^{O} - \boldsymbol{x}_{ij}^{O})^{2}
    \label{eq:train_autoencoder}
\end{align}
where $n$ is the training batch size.

\textbf{Text component}. We utilize the text component to read and understand words in image ads.   An OCR model is first applied to detect and recognize $M$ words, $\{w_{i1}, w_{i2}, ... w_{iM}\}$ from ads. Then, these words will be embedded into word vectors $\{ \boldsymbol{x}_{i1}^{W}, \boldsymbol{x}_{i2}^{W}, ..., \boldsymbol{x}_{iM}^{W} \}$ by using FastText. To be noticed, FastText embedding is necessary in this stage. Our empirical experiences show that even with the most advanced open-source OCR tool, there are still many words can not be detected or recognized correctly. It is often missing some characters in words. Since FastText can embed out-of-vocabulary (OOV) words, it can eliminate the recognition issues a lot. Third, these word embeddings will be input to a sequence model (e.g. BLSTM) to learn shared word representations $\{ \boldsymbol{z}_{i1}^{O}, \boldsymbol{z}_{i2}^{O}, ..., \boldsymbol{z}_{iM}^{O} \}$. Formal formulations are shown as follows:
\begin{gather}
    \{w_{ij}\}_{j=1}^{M} = f_{ocr}(\text{IMAGE}_i ; \theta_{\text{fixed}})\\
    \boldsymbol{x}_{ij}^{W} = f_{FastText}(w_{ij} ; \theta_{\text{fixed}})\\
    \boldsymbol{z}_{ij}^{W} = f_{BLSTM}(\boldsymbol{x}_{ij}^{W} ; \theta)
    \label{eq:textcomponent}
\end{gather}

\textbf{Discussion}. We design a shared bottom module to extract and learn general multimodal features for the next two tasks. We highlight the advantage of this module in four folds: 1) Understanding ads needs not only visual information but also text information; 2) The \textbf{\textit{rhetoric}} issue is initially addressed by an autoencoder with reconstruction loss; 3) Some techniques we select such as FasText improves the robustness of the representation; 4) Since this is a shared module, the number of parameters we need to learn decreases a lot which also prevents the overfitting.

\subsection{Hierarchical Multimodal Attention Module}

\begin{align*}
    \text{\textbf{Input}:} \{(\boldsymbol{z}_{i}^{V}, {\{\boldsymbol{z}_{ij}^{O}\}}_{j=1}^{L}, {\{\boldsymbol{z}_{ij}^{W}\}}_{j=1}^{M})\} \rightarrow \text{\textbf{Output}:} \{\boldsymbol{r}_{i}^{T}, \boldsymbol{r}_{i}^{S}\}
\end{align*}
In the second phrase, all features obtained from the first phrase are fused into a task-specific feature vector by the hierarchical multimodal attention module (HMAM). Specifically, topic and sentiment task will be processed by two HMAMs, respectively.  We design two attention mechanisms to form this module: \textbf{intro-modality attention} and \textbf{inter-modality attention}.

\textbf{Intro-modality attention}. This attention reads all feature vectors, extracted by the first phrase, from the same modality, and generate linear weights to fuse them. Since visual modality, $\boldsymbol{z}_{i}^{V}$, only contains one feature vector, we do not process it in the first attention. For the other object and text modality features, we apply intro-modality attention, respectively. Let ${\{\boldsymbol{z}_{ij}\}}_{j=1}^{N}$ be the feature vectors, where $N$ is the number of vectors in this modality. An attention block first filters them with a kernel $\boldsymbol{q}$ via dot product, yielding a set of corresponding significance ${\{p_{ij}\}}_{j=1}^{N}$. They are then passed to a softmax function to generate positive weights $\{a_{ij}\}$  with $\sum_{j=1}^{N}a_{ij} = 1$. These two operations takes the following mathematical form, respectively:
\begin{gather}
    p_{ij} = \boldsymbol{q}^{\mathsf{T}} \boldsymbol{z}_{ij}
    \label{eq:attention_kernal}
\end{gather}
\begin{gather}
    a_{ij} = \frac{\text{exp}(p_{ij})}{\sum_{j'=1}^N \text{exp}(p_{ij'})}
    \label{eq:attention_softmax}
\end{gather}
The final representation for one modality is calculated by the equation:
\begin{align}
    \boldsymbol{m}_{i} = \sum_{j=1}^{N}a_{ij}\boldsymbol{z}_{ij}
    \label{eq:attention_sum}
\end{align}
This intro-modality attention simulates human visual system. It perceives more important information by weighting them more and aggregate all features into one feature vector to represent one modality. Besides, from equation (\ref{eq:attention_kernal}) we can easily deduce that this attention can take any numbers of feature vectors as long as the length of these vectors keep same. This increase the flexibility of our framework. Moreover, two $\boldsymbol{q}$ for two modalities can be trained by standard back-propagation and gradient descent.

\textbf{Inter-modality attention}. This attention weights three modality feature vectors and concatenates them into one task-specific representation. Since we only have three modalities, using the kernel method same with the intro-modality attention accounts for overfitting. Inspired by annealing algorithm, we simplify the attention mechanism for inter-modality. First, we initialize the attention score vector ${\{a_{i}\}}^{3}_{i=1}$ directly. Then, for each modality representation, we use $a_i$ to weight them and get final task-specific representation by following equation:
\begin{align}
    \boldsymbol{r}_{i} = \begin{bmatrix}
    a_1\\
    a_2\\
    a_3
    \end{bmatrix} \cdot \begin{bmatrix}
    \boldsymbol{z}_{i}^{V}&\boldsymbol{m}_{i}^{O}&\boldsymbol{m}_{i}^{W}
    \end{bmatrix}
    \label{eq:simple_attention_weight}
\end{align}
By the inter-modality attention, the final feature vector is constructed and passed to the next phrase for prediction. The attention score vector can be also trained by back-propagation and gradient descent with specific settings. But different from the intro-modality attention, the inter-modality attention takes fewer parameters which prevent the overfitting issue. Meanwhile, we do not use softmax function to restrict attention weights to be positive. The reason is that people adopt many design tricks to make ads more attractive, which has no contribution even negative contribution to the final prediction (This is why for one ad image, different annotators label them differently).

\textbf{Discussion}. We process multimodality information in which there are multiple feature vectors in an hierarchical manner. In the first-level attention, feature vectors within one modality are weighted and aggregated together. Different attention filter kernels are initialized and trained for different modality. In the second-level attention, three modality feature vectors are scored by a simplified attention vector then concatenated as the final task representation. The attention can prevent overfitting and some scores may be trained to negative to penalize misleading information.

\subsection{Multitask Prediction Module}

\begin{align*}
    \text{\textbf{Input}:} \{\boldsymbol{r}_{i}^{T}, \boldsymbol{r}_{i}^{S}\} \rightarrow \text{\textbf{Output}:} \{(\boldsymbol{y}^{T}_{i}, \boldsymbol{y}^{S}_{i})\}
\end{align*}
Multitask prediction module contains two prediction modules which are applied to topic and sentiment representations from last phrase, respectively. These representations are first transformed into a low-dimension space with the following equation:
\begin{align}
    \boldsymbol{r}^{0}_{i} = f_{relu}(\boldsymbol{r}; \theta)
    \label{eq:relu}
\end{align}
where $f$ is the non-linear activation function, $RELU$ and $\theta$ is the parameters can be trained. Then, the $\boldsymbol{r}^{0}$ is fed into the output layer with sigmoid activation function that can process multiple possible labels for one sample that are not mutually exclusive:
\begin{gather}
    \boldsymbol{y'}_{i} = \boldsymbol{w}^{\mathsf{T}}{\boldsymbol{r}^{0}_{i}} + \boldsymbol{b}\\
    \boldsymbol{y}_{i} = \begin{bmatrix}
    \frac{1}{1+\text{exp}(-{y'}_{i}^{0})}\\
    \frac{1}{1+\text{exp}(-{y'}_{i}^{1})}\\
    \vdots\\
    \frac{1}{1+\text{exp}(-{y'}_{i}^{P})}
    \end{bmatrix}
\end{gather}
where $P$ is the number of total labels in one task.

\textbf{Discussion}. With the help of sigmoid, the neural network models the probability of a class $y_i$ as Bernoulli distribution. The probabilities of each class is independent from the other class probabilities, so we can use the threshold 0.5 as usual to do prediction (i.e., if the $i$th  probability is larger than 0.5, the network will output the $i$th label.

\subsection{Training Methodology}

We construct a multitask loss to train the proposed framework. It consists of three losses: rhetoric loss, topic loss and sentiment loss. Rhetoric loss, as described in equation (\ref{eq:train_autoencoder}), aims to distinguish the true meaning or metaphor meaning of an object in ads. Since there is no supervised information, this part is trained in an unsupervised manner. The topic and sentiment losses (i.e., $\mathcal{L}_{topic}$ and $\mathcal{L}_{sentiment}$) have same mathematical form which can be called multi-label loss $\mathcal{L}_{ml}$. It can be calculated as follows:
\begin{align}
    \mathcal{L}_{ml} = - \sum_{i=1}^{N}\sum_{j=1}^{P}y_{i}^{j}\text{log}{\hat{y}}_{i}^{j} + (1-y_{i}^{j})\text{log}(1-{\hat{y}}_i^{j})
    \label{eq:topic_sentiment_loss}
\end{align}
Here, $y_{i}^{j}$ denote the ground-truth of $i$-th ad on the $j$-th label. $y_{i}^{j}=1$ if $j$-th label is the relevant label, otherwise $y_{i}^{j}=0$. ${\hat{y}}_{i}^{j}$ is the prediction output. $P$ is the number of total label in one task and $N$ is the batch size.

The overall loss function turns out to be the following form:
\begin{align}
    \mathcal{L}_{multitask} = \mathcal{L}_{share} + \alpha\mathcal{L}_{topic} + \beta {L}_{sentiment}
    \label{eq:multitaskloss}
\end{align}
where $\alpha$ and $\beta$ are the balance coefficients which control the interaction of the loss terms.

\section{Experiment}
\label{sec:exp}

In this section, we first provide a detailed description of the dataset and some evaluation metrics used in the experiments. Then we design a set of experiments to evaluate the performance of the proposed multimodal multitask framework. Furthermore, we conduct ablation study to demonstrate the effectiveness of designed modules

\subsection{Dataset}

%

We evaluate our framework in a latest ads dataset released from \cite{hussain2017automatic}. There are 64832 image ads with 38 topic labels and 30 sentiment labels in the dataset. One image may annotated by multiple topic or sentiment labels. Since some of them only includes topic labels or sentiment labels, to verify our framework, we filter the dataset and only use a sub-dataset where every ad image annotated in both topic and sentiment. The sub-dataset contains 30,000 images which is still a large enough to train a deep learning model. After we apply a OCR technique comes from google, we observe that more than 67\% ads contains text information which prove the necessary of using multimodal learning. Unless otherwise stated, we use 70\% of the dataset to train, 10\% to validate and left 20\% to test the proposed framework.


\subsection{Compared Approaches}
Since both topic and sentiment tasks can be cast as multilabel classification, we select ResNet-50 and ResNet-101 with last activation function replaced by $sigmoid$ as baselines. Then we compare our framework works to some state-of-the-art (SOTA) multilabel classification framework for images. Their details are as follows:

\begin{itemize}
    \item \textbf{ResNet} \cite{he2016deep}: a very competitive image classification model which has been applied to a lot of areas. We replace its last activation function by $sigmoid$ and train it with binary cross-entropy loss function. We fine-tune last layer or all layers of ResNet-50 and ResNet-101 models trained on ImageNet \cite{deng2009imagenet} on ad image dataset as baseline.

    \item \textbf{C2AE} \cite{yeh2017learning}: \textbf{C}anonical \textbf{C}orrelated \textbf{A}uto\textbf{E}ncoder, a deep learning based multilabel image recognition framework. It learns to embed features and labels jointly which relates feature and label so to improve recognition.

    \item \textbf{GCN} \cite{chen2019multi}: a recently proposed model based on graph convolutional networks (GCNs). It models label dependencies by constructing graphs and applies GCN.
\end{itemize}
Following conventional settings from \cite{wang2016cnn,yeh2017learning,chen2019multi}, we report the average per-class F1 (F1-C) and
the average overall F1 (F1-O) for performance evaluation.

\subsection{Implementation Details}
We build the framework by PyTorch 1.0 \cite{paszke2017automatic}. Unless otherwise noted, our configuration is as follows: 1) In the shared bottom module, The image-level features are obtained by average pooling 2048\textit{D} features from the \textit{res-5c} block of a pre-trained ResNet-152. Image-level features will be fed into a 2-layer MLP to learn a 1024\textit{D} shared global features. The object-level features are extracted from the \textit{fc6} layer of an improved Faster-RCNN model \cite{ren2015faster} trained on the Visual Genome \cite{krishna2017visual} objects and attributes as provided in \cite{anderson2018bottom}. These features will be passed to an autoencoder and we use the 1024\textit{D} latent features as the shared object-level features. For OCR, we use an open-source tool named Tesseract \cite{smith2007overview}, comes from google. Then FastText embedding will embed them into 300\textit{D} vectors. A 1-layer BLSTM with hidden-size 512 will further process these word vectors to 1024\textit{D} vectors.

We train our framework in an end-to-end manner. Adamax \cite{kingma2014adam} with learning rate 0.001, momentum 0.9 and weight decay 0.0001 is applied to optimize the network. Every 15 epochs, we multiply the learning rate by 0.1. The balance coefficient $\alpha$ and $\beta$ is set to 200 and 50, respectively.

\subsection{Experimental Result}
We presents our evaluation results from three perspectives. We first compare our models to other STOA multilabel classification baselines. We then demonstrate the parameter sensitivity of the proposed framework. We end this section by conducting ablation study to verify the effectiveness of each module.

\begin{table}[]
\centering
\caption{Quantitative comparison for topic and sentiment prediction with different methods. We apply all baselines with their official open-sourced code and benchmark them on the ad dataset. For both two tasks under all metrics, our framework achieves the best performance. The overall performance for topic prediction is much better and we own this to our multimodal learning based design.}
\label{tab:topic_sentiment_all}
\begin{adjustbox}{width=\columnwidth,center}
\begin{tabular}{@{}lcccccc@{}}
\toprule
\multicolumn{1}{c}{\multirow{2}{*}{Method}} & \multicolumn{3}{c}{Topic (All)}                  & \multicolumn{3}{c}{Sentiment (All)}                          \\ \cmidrule(l){2-7}
\multicolumn{1}{c}{}                        & mAP            & F1-C           & F1-O           & mAP                        & F1-C           & F1-O           \\ \cmidrule(r){1-7}
ResNet50 (Last)                             & 0.151          & 0.082          & 0.257          & 0.242                      & 0.136          & 0.376          \\
ResNet50 (All)                              & 0.206          & 0.130          & 0.363          & 0.261                      & 0.153          & 0.400          \\
ResNet101 (Last)                            & 0.157          & 0.088          & 0.271          & \multicolumn{1}{c}{0.244} & 0.138          & 0.384          \\
ResNet101 (All)                             & 0.215          & 0.138          & 0.379          & 0.265                      & 0.162          & 0.408          \\
C2AE                                        & -              & 0.146          & 0.314          & -                          & 0.201          & 0.422          \\
GCN                                         & 0.120          & 0.051          & 0.183          & 0.223                      & 0.110          & 0.339          \\ \cmidrule(r){1-7}
Ours-Single-task                            & 0.290          & 0.214          & 0.482          & 0.284                      & 0.192          & 0.439          \\
\textbf{Ours-Multi-task}                    & \textbf{0.382} & \textbf{0.371} & \textbf{0.585} & \textbf{0.292}             & \textbf{0.216} & \textbf{0.453} \\ \bottomrule
\end{tabular}
\end{adjustbox}
\end{table}

\subsubsection{Comparison with Other Approaches.} Table \ref{tab:topic_sentiment_all} presents both topic and sentiment results of each model. Deep$M^2$AD performs the best and significantly outperforms all baseline models on all evaluation metrics, which showcases the effectiveness of the proposed method on ads understanding. For more in-depth performance analysis, we realize that even a very simple $MLP$ with multimodality information, the performance increases a lot which show us the direction of images ads understanding. We should pay more attention on multimodal information extraction and fusion rather than designing deeper networks such as ResNet, DenseNet and so on. We also observe that the performance of topic prediction increases more than the performance of sentiment prediction. We attribute this to two reasons: 1) The text in image ads exposes more information to our framework and leads better topic results (e.g., some text may show the topic directly); 2) From the mAP column of sentiment results, we note that only using resnet can get a comparable results which means that the human emotion for ads depends on the whole images more. In other words, the hierarchical multimodal attention module for objects, words and modalities is more effective on topic prediction.

\begin{figure}[!ht]
\centering
\begin{subfigure}{1.0\columnwidth}
  \centering
  \includegraphics[width=1.0\columnwidth]{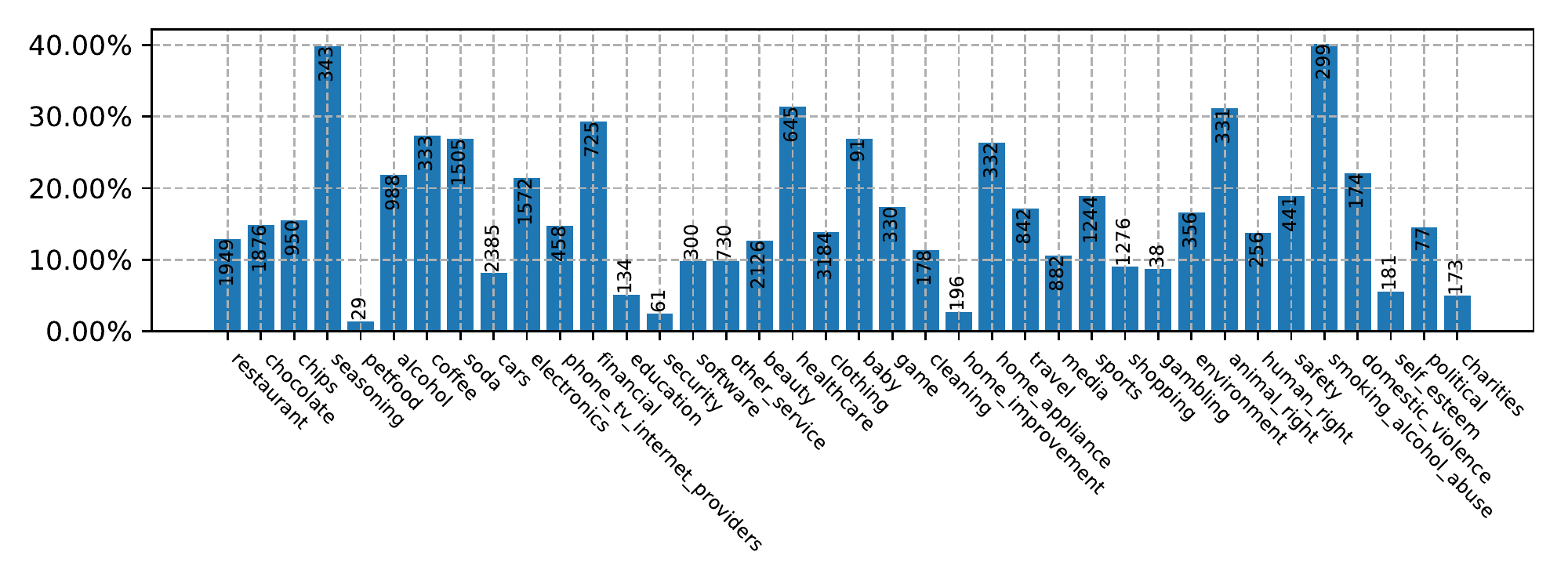}
    \caption{The improvements over each class for the topic task.}
  \label{fig:ap_improve_topic}
\end{subfigure}
\begin{subfigure}{1.0\columnwidth}
  \centering
  \includegraphics[width=1.0\columnwidth]{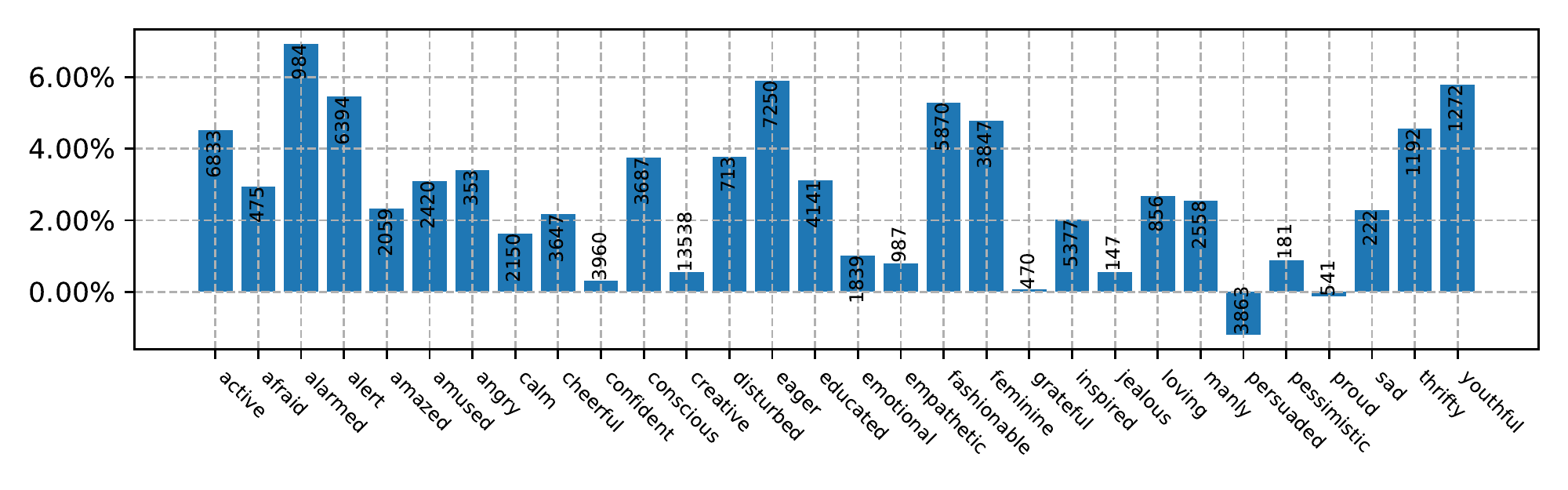}
  \caption{The improvements over each class for the sentiment task.}
  \label{fig:ap_improve_sentiment}
\end{subfigure}
\caption{The improvements of our proposed framework over each class on two ad understanding tasks. The value attached to each bar represents the number of the corresponding label in the test set. Within the imbalanced dataset, our model performs significant improvement over almost all classes in both tasks.}
\label{fig:ap_improvement_both}
\end{figure}

The improvements over each class on two tasks are shown in Figure \ref{fig:ap_improve_topic} and \ref{fig:ap_improve_sentiment}. As it shows, the average precision of all classes in topic task are improved even the number of ads in the one of classes is very small (e.g., the petfood class only contains 29 images). This demonstrates that our framework can mitigate the data imbalance issue. We also notice that the larger improvements are not occurred in classes with larger samples. This indicates that: 1) our framework do not depend on more modality information and training samples; 2) the visual \textit{rhetoric} can be classified with our unsupervised training because the top3 improvement classes are ``seasoning", ``smoking alcohol abuse" and `` animal right" which are use \textit{rhetoric} design widely. For the sentiment prediction, almost all classes are improved except two are decreased by about 1\%. The two classes are ``persuaded'' and ``proud''. This may be attributed to the misleading of multitask training (i.e., our framework pays more attention on local features for better task prediction which may produce error signals to sentiment prediction in very rare samples).

\begin{figure}[!ht]
\centering
\begin{subfigure}{.5\columnwidth}
  \centering
  \includegraphics[width=1.0\linewidth]{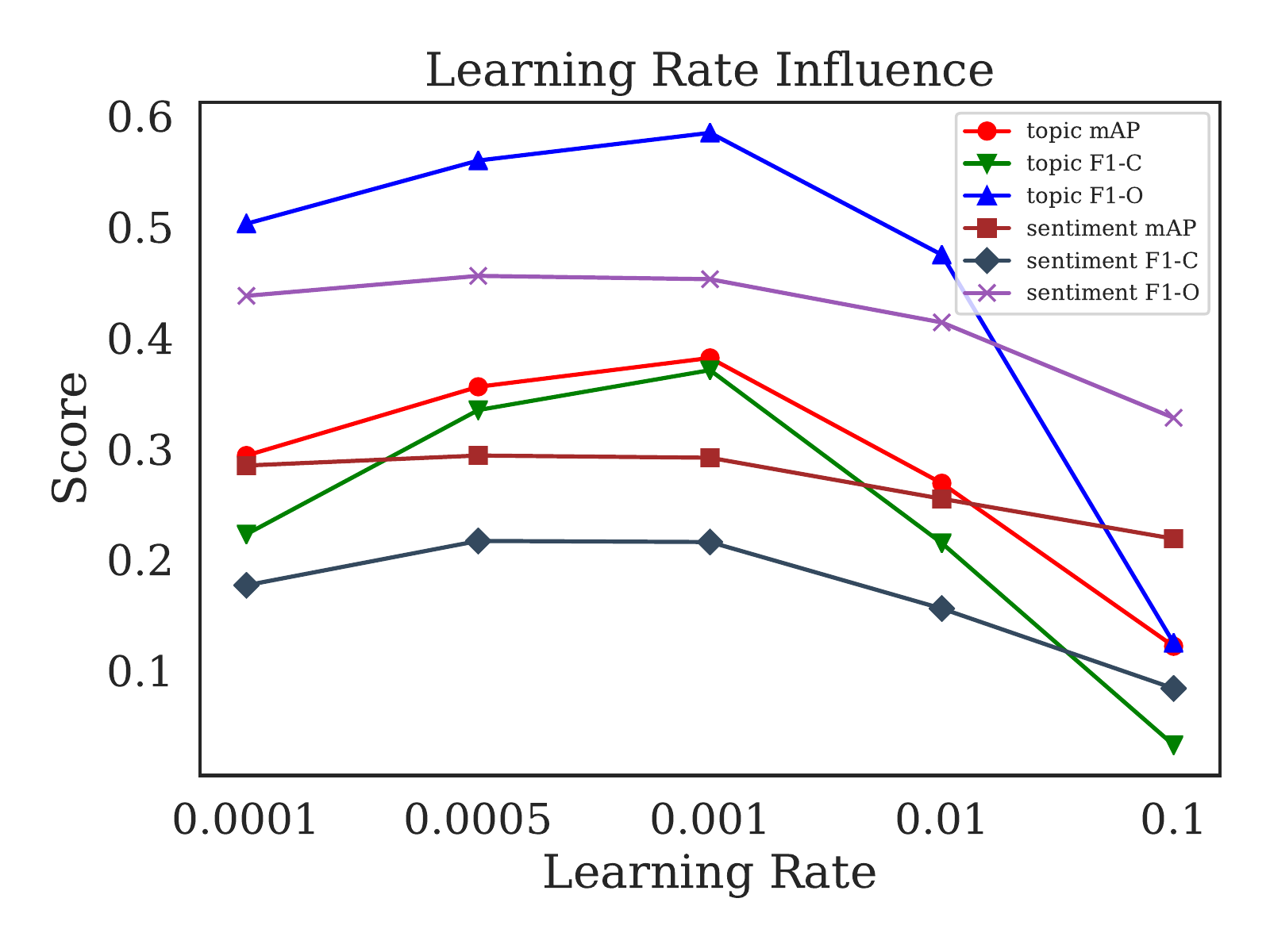}
  \caption{}
  \label{fig:lr_influ}
\end{subfigure}%
\begin{subfigure}{.5\columnwidth}
  \centering
  \includegraphics[width=1.0\linewidth]{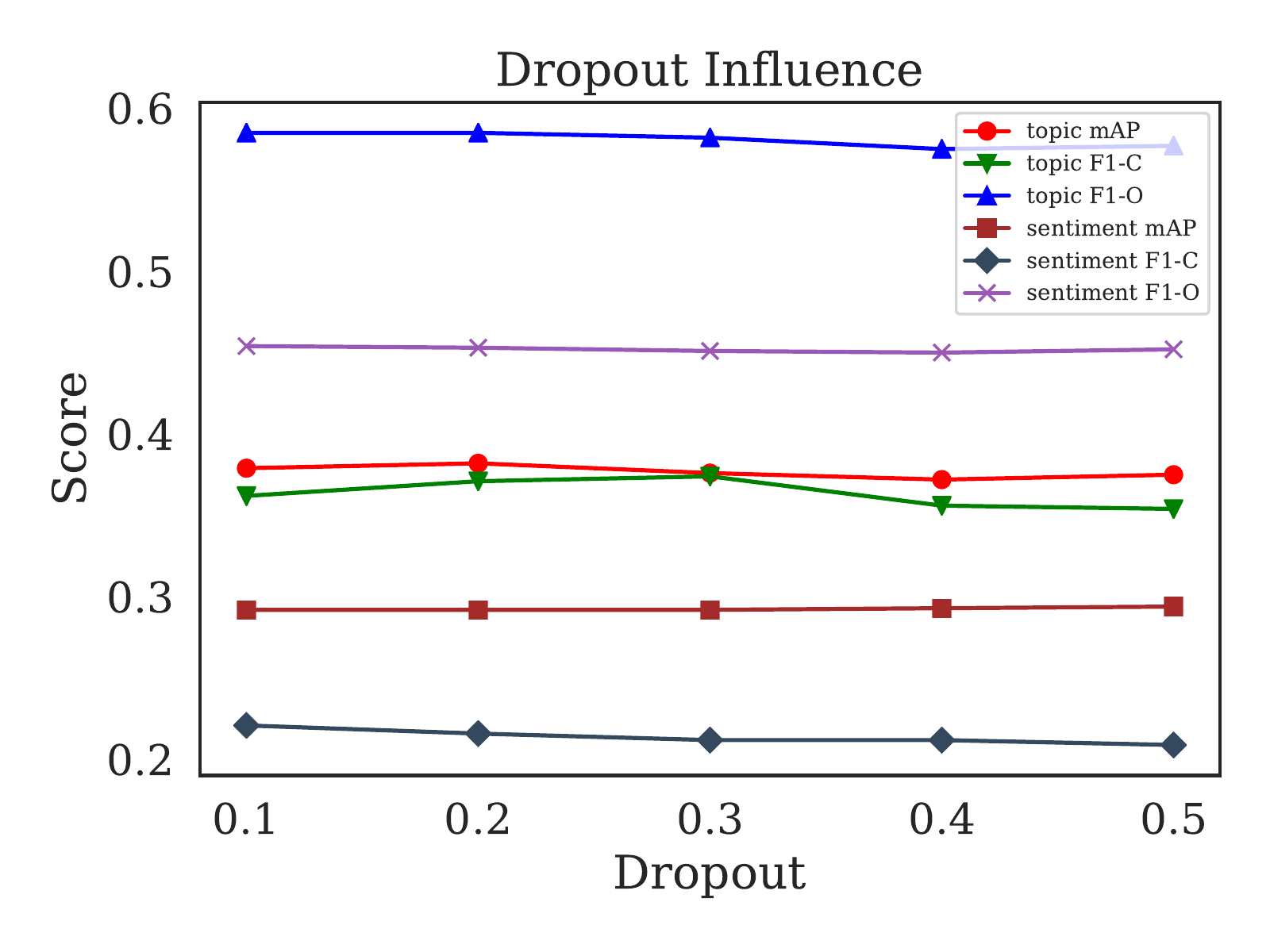}
  \caption{}
  \label{fig:dropout_influ}
\end{subfigure}
\caption{The influence of different hyper-parameter settings. The left figure shows the trend within learning rate ranging from 0.0001 to 0.1. Our framework work is quite stable under a large range which indicates it is not very sensitive to learning rate. The right figure illustrates the dropout influence. Since dropout is used to prevent overfitting, our framework is not influenced by it which proves the model does not overfit on the dataset.}
\label{fig:parainflu}
\end{figure}

\subsubsection{Parameter Sensitivity Analysis.}

\begin{figure}
\begin{subfigure}{.5\columnwidth}
  \centering
  \includegraphics[width=1.0\linewidth]{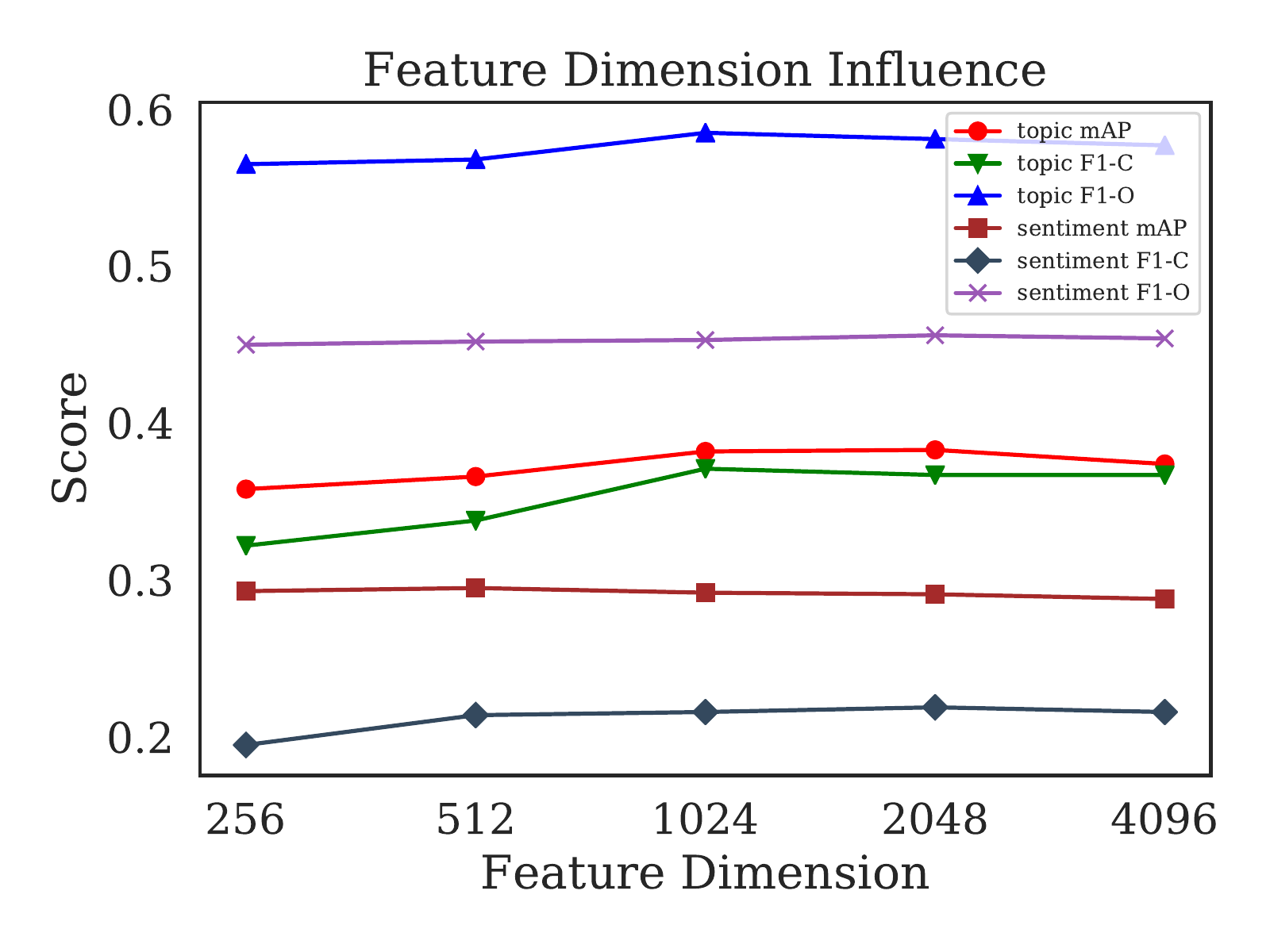}
  \caption{}
  \label{fig:feature_dimension_influ}
\end{subfigure}%
\begin{subfigure}{.5\columnwidth}
  \centering
  \includegraphics[width=1.0\linewidth]{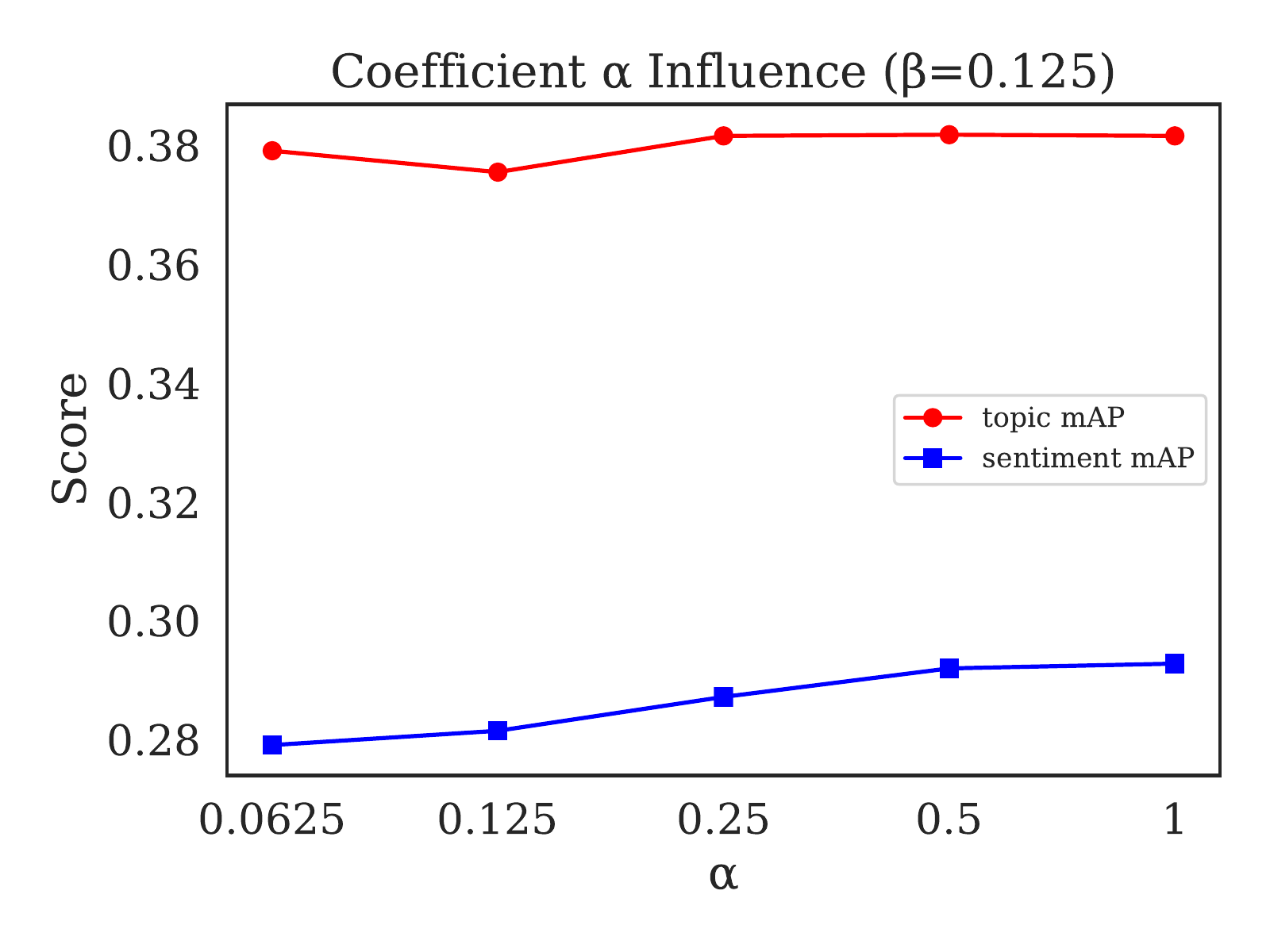}
  \caption{}
  \label{fig:coefficient_influ}
\end{subfigure}
\caption{The influence of different framework settings. The left is the influence of different feature dimensions. The figure illustrates that when the dimension increases, the performance first increases and then decreases. This is because fewer dimensions will lead to information lost and more dimensions would introduce noise. The right is the coefficient influence. We fix $\beta$ and tune $\alpha$. When the $\alpha$ becomes larger, both the topic and sentiment prediction performance first becomes better and then steady.}
\label{fig:feat_co_influ}
\end{figure}

We present the test results under different parameters to verify the robustness of our framework. Figure \ref{fig:lr_influ} illustrates the influence of learning rate. The best performance is achieved at 0.001. In general, the performance is pretty stable under small learning rates. Dropout is to alleviate the overfitting of the network. We set dropout between layers and as Figure \ref{fig:dropout_influ} shows, our framework performs well within all dropout settings. This proves that we do not over-fit the model to the dataset.

The dimension of multimodal features is also an important hyper-parameter. After processed by shared bottom module, these features have same dimension numbers. We set different dimension values to check its influence as shown in Figure \ref{fig:feature_dimension_influ}. The mAP is first increases then drops. The reason is that too small dimension accounts for the information lost and too large dimension may adopt noise to the feature vectors.

The influences of balance coefficients $\alpha$ and $\beta$ are demonstrated in Figure \ref{fig:coefficient_influ}. We fix one and adjust the other to check the influence. As $\alpha$ controls the topic loss, it is not surprised the topic performance is increased as the $\alpha$ value increased. And as the topic performance increased, the sentiment performance is improved which proves the effectiveness of our multitask loss.

\subsubsection{Ablation Study}
We conduct detailed ablation study to verify the effectiveness of each module in our framework. 1) Rhetoric autoencoder component performs better on topic prediction than sentiment prediction. This meets our expectation since this component only is only utilized by object-level features which contribute topic more. 2) Similar to 1), the hierarchical attention module performs remarkably on topic task. The reason is this module mimic human visual attention so that it can extract import local features. Meanwhile, the sentiment task is influenced by global features more. 3) Multitask loss improves the sentiment prediction a lot. We own this to the relevance between topic and sentiment. From the first two components, we already have a better topic representation which can assist sentiment representation.

\begin{table}[]
\centering
\caption{Ablation Study. We keep other parts fixed and remove the specific component to verify it. The first two components performs better on topic task and the last multitask loss is better for sentiment task. The former is beneficial for understanding region features of ads and the latter indicates topic understanding promotes sentiment understanding.}
\label{tab:Ablation_Study}
\begin{adjustbox}{width=\columnwidth,center}
\begin{tabular}{@{}lcccccc@{}}
\toprule
\multicolumn{1}{c}{\multirow{2}{*}{Remove Modules}} & \multicolumn{3}{c}{Topic} & \multicolumn{3}{c}{Sentiment} \\ \cmidrule(l){2-7}
\multicolumn{1}{c}{} & mAP & F1-C & F1-O & mAP & F1-C & F1-O \\ \midrule
- Rhetoric AutoEncoder & -1.77\% & -5.20\% & -2.50\% & -0.82\% & -1.50\% & -1.30\% \\ \midrule
- Hierarchical Attention & -2.92\% & -6.00\% & -3.30\% & +0.36\% & -1.00\% & -0.30\% \\ \midrule
- Multitask Loss & -0.65\% & -1.40\% & -0.70\% & -0.36\% & -4.50\% & -2.30\% \\ \bottomrule
\end{tabular}
\end{adjustbox}
\end{table}

\section{Related Work}
In this section, we summarize some closely related works on ads analysis. Ads related research has a long history. Some early works focus on predicting click-through rates in ads using low-level visual features \cite{azimi2012visual,cheng2012multimedia}. In \cite{mcduff2014predicting,okada2018advertisement}, some approaches are developed to predict how much human viewers will like an ad by capturing their facial expressions. The user affect and saliency are utilized in \cite{mei2012imagesense,yadati2013cavva} to determine the best placement of a commercial in a video stream, or of image ads in a part of an image. \cite{sanchez2002shot,gauch2006finding} proposed to detect whether the current video shown on TV is a commercial or not, and a solution for detecting human trafficking advertisements is provided in \cite{sethi2013large}. A method for extracting the object being advertised from commercials (videos) is proposed in \cite{zhao2011discovering}, by looking for recurring patterns (e.g. logos). Human facial reactions, ad placement and recognition, and detecting logos, are quite distinct from our goal of understanding the messages of ads. Although lots of efforts have been devoted to ads analysis, uncovering the meaning of ads attracts little attention. One of main reasons is the lack of dataset, and this issue is tackled by \cite{hussain2017automatic}, where the authors make great efforts to collect and propose a new dataset for image and video ads understanding. In \cite{hussain2017automatic}, some baselines are presented for each single prediction task, while our framework unifies the topic and sentiment prediction.

\section{Conclusion and Future Work}


In this paper, we present a novel deep learning based framework to simultaneously predict the topic and sentiment of advertisement. Our model is able to effectively integrate the different multimodal information contained in the ad for joint topic and sentiment prediction, and the feature importance is well exploited using the proposed hierarchical multimodal attention module. Extensive experiments are conducted on a recently proposed dataset, and we provide a benchmark for comparison. Compare to other related approaches, our framework improves the performance on both prediction tasks significantly. The ablation study shows the effectiveness of our proposed modules. In the future, we plan to adopt neural architecture search (NAS) techniques to further improve the performance.


\newpage
\bibliography{ads_multimodal}
\bibliographystyle{aaai}

\end{document}